# Joint Resource Optimization, Computation Offloading and Resource Slicing for Multi-Edge Traffic-Cognitive Networks

Ting Xiaoyang, Minfeng Zhang, Shu gonglee, Saimin Chen Zhang*
School of Information and Electrical Engineering, Shaoxing University Yuanpei College,
Shaoxing, Zhejiang, 312000, China, *corresponding author{Saimin.cz@ypc.edu.cn}

**Abstract**
The evolving landscape of edge computing envisions platforms operating as dynamic intermediaries between application providers and edge servers (ESs), where task offloading is coupled with payments for computational services. Ensuring efficient resource utilization and meeting stringent Quality of Service (QoS) requirements necessitates incentivizing ESs while optimizing the platform's operational objectives. This paper investigates a multi-agent system where both the platform and ESs are self-interested entities, addressing the joint optimization of revenue maximization, resource allocation, and task offloading. We propose a novel Stackelberg game-based framework to model interactions between stakeholders and solve the optimization problem using a Bayesian Optimization-based centralized algorithm. Recognizing practical challenges in information collection due to privacy concerns, we further design a decentralized solution leveraging neural network optimization and a privacy-preserving information exchange protocol. Extensive numerical evaluations demonstrate the effectiveness of the proposed mechanisms in achieving superior performance compared to existing baselines.

**Keywords--**Mobile edge computing; computation offloading; resource slicing; DRL-driven traffic prediction

## I. Introduction

In recent years, a surge of novel applications, such as augmented reality, interactive gaming, and autonomous driving, has placed unprecedented demands on computational and network resources. These applications are both resource-intensive and delay-sensitive, necessitating robust and low-latency computing frameworks. Multi-access edge computing (MEC), previously referred to as mobile edge computing, has emerged as a promising paradigm to address these challenges. By bringing storage, computing, and networking resources closer to end-users at the network edge, MEC enables low-latency services while enhancing privacy and security by processing data locally.

MEC relies on a distributed infrastructure of computing nodes, often termed edge servers (ESs), to handle users' tasks. Unlike centralized cloud computing with abundant resources, edge servers are typically resource-constrained, posing challenges in meeting the stringent Quality of Service (QoS) requirements of modern applications. Although the deployment of edge resources continues to expand, their ad hoc configuration often limits scalability and accessibility, as many applications utilize these resources privately. To address these limitations, cooperative edge computing has been proposed, enabling collaborative resource sharing among edge servers to meet peak computational demands. For instance, offloading tasks to non-nearest edge servers within the same metropolitan area network has been shown to significantly enhance users' quality of experience.

Building on these developments, this paper envisions the evolution of MEC systems into platforms resembling crowdsourcing ecosystems. Crowdsourcing, defined as obtaining services, ideas, or resources through contributions from a large group of individuals, offers a scalable and flexible approach to problem-solving. Integrating crowdsourcing principles into MEC systems, referred to as crowdsourcing-like MEC, provides a compelling vision for achieving edge-as-a-service. In such systems, edge service providers with idle resources can join the platform to generate additional revenue, enhancing resource utilization efficiency. Simultaneously, application providers can access elastic and scalable edge computing services tailored to their needs.

To fully realize a crowdsourcing-like edge computing system, several critical challenges must be addressed. These challenges arise from the inherent complexity of managing self-interested participants, ensuring security, adhering to policy compliance, and achieving effective system management. Among these, we focus on the following key issues:

1- Quality of Service (QoS) Assurance: How can the system guarantee the QoS requirements of applications? Application providers will only offload their tasks if the platform can reliably meet these stringent requirements.
2- Resource Allocation: How should edge servers (ESs) allocate resources to applications? Over-provisioning resources can enhance service quality but may result in reduced revenue for ESs due to the fixed rewards from the platform and resource costs.
3- Task Dispatching: How can tasks be optimally dispatched to multiple ESs hosting the same service to maximize system performance?
4- Reward Design: What reward mechanisms should the platform implement to fairly compensate ESs for their contributions while accounting for the self-interest of both the platform and ESs, as both aim to maximize their revenue?
5- Privacy and Security: How can the system function effectively when participants, such as application providers and ESs, refuse to share sensitive information (e.g., task details or resource capacities) due to privacy concerns?

These challenges are intricately interrelated, and addressing them in isolation risks rendering the system either infeasible or suboptimal. A holistic approach is thus imperative to develop effective solutions that balance performance, fairness, and practicality.

In this paper, we consider a scenario where application providers make upfront payments to the platform before task offloading begins. Our objective is to address the aforementioned challenges and design efficient and practical mechanisms that ensure participation from all stakeholders—application providers, edge servers (ESs), and the platform—while meeting their diverse requirements for consuming or providing edge services. Specifically, we establish a game-theoretic framework to model the interactions among the three entities and propose both centralized and decentralized solutions, depending on the level of information the platform can access.

Our key contributions are summarized as follows:

- **Comprehensive System Modeling:**
  We present a heterogeneous crowdsourcing-like multi-server and multi-application MEC system, where edge servers have varying resource capacities, costs, and revenue expectations, while applications differ in workload, budget, and QoS requirements (see Fig. 1). We formulate a joint optimization problem for revenue maximization, resource allocation, and task offloading, capturing the system's diverse constraints and interactions.
- **Game-Theoretic Centralized Solution:**
  We model the interactions between the platform and ESs as a Stackelberg game and propose a novel centralized solution based on Bayesian optimization. Unlike traditional backward induction approaches, we develop an efficient algorithm to maximize the platform's revenue under given resource allocations at ESs. The optimal policies for task offloading, resource allocation, and reward distribution are derived using this technique.
- **Privacy-Preserving Decentralized Mechanism:**
  To address scenarios where application providers and ESs are unwilling to share sensitive information (e.g., task details, resource capacities), we design a decentralized solution using neural network optimization. This mechanism allows application providers to submit aggregated workload information, safeguarding their privacy, while ESs determine resource allocations locally without exposing private details.

- **Performance Evaluation and Insights:**
  We conduct extensive numerical evaluations to compare the proposed centralized and decentralized mechanisms with representative baselines. Results demonstrate that both approaches significantly improve the platform's revenue and incentivize ESs to contribute resources. Interestingly, the decentralized mechanism achieves even better revenue for individual ESs compared to the centralized mechanism, offering valuable insights into designing and operating crowdsourcing-like MEC systems.

  The remainder of this paper is organized as follows. Section 2 discusses related works, Section 3 introduces the system model and problem formulation. Section 4 presents the optimization-based centralized mechanism. Section 5 elaborates on the DRL-based decentralized solution. Numerical evaluations and results are discussed in Section 6. and the paper concludes in Section 7.

## II. RELATED WORKS

Edge computing has evolved from a proximity-based extension of cloud services into an intelligent, service-centric ecosystem that must jointly manage radio resources, computing capacity, and service-level constraints under time-varying user demands [1]-[4]. This evolution has motivated a rich body of work on (i) resource-aware task offloading, (ii) digital-twin (DT) empowered decision making, (iii) learning-based optimization (especially reinforcement learning), and (iv) distributed coordination mechanisms such as federated learning (FL), sometimes reinforced by blockchain and incentive design. In parallel, earlier research on dynamic routing and energy-aware forwarding in mobile and sensor networks has highlighted how fast-changing conditions require adaptive decision rules, rather than static heuristics, which conceptually aligns with today's adaptive edge control and traffic-aware optimization (e.g., optimum path discovery in MANETs and opportunistic routing for power reduction in WSNs) [5]-[8].

### *A. Task Offloading and Resource Allocation in Edge Computing:*

A central theme in mobile edge computing (MEC) is how to select offloading targets and allocate scarce edge resources while meeting latency or reliability requirements. Recent works have explored different physical-layer and network-layer knobs to improve task completion performance [9], [10]. For example, IRS-assisted MEC has been proposed to improve QoS under delay constraints and challenging channel conditions, demonstrating how radio environment control can reshape feasible offloading regions and service guarantees [11]. More broadly, several studies show that effective offloading is rarely a single-layer decision; it requires jointly handling communication rate, computation scheduling, and service constraints, which becomes more complex as applications diversify and as the number of edge nodes grows. This need for joint decision making also appears in multi-layer optimization frameworks for QoS-aware resource allocation in cellular HetNets, including NOMA-based heterogeneous deployments where the coupling between tiers and user classes makes single-layer tuning insufficient [12].

Beyond radio enhancements, another influential line of work studies how to optimize service satisfaction and timeliness metrics at the edge. In particular, Age of Information (AoI) has become an important concept for systems where data freshness matters. AoI-aware service satisfaction optimization has been investigated for digital-twin empowered edge computing, demonstrating that user satisfaction can depend on more than raw delay, it also depends on how quickly the system updates and reflects the physical process state [13]. Complementing this, AoI-aware query services in DT-empowered MEC emphasize how query response quality can be coupled to timeliness and edge-side decisions, making the offloading and service pipeline more than a compute scheduling problem [14]. These studies support broader systems view: performance must be judged by end-to-end service metrics, not only by isolated link or CPU metrics. A related QoS-driven viewpoint also appears in live streaming over edge-cloud aided HetNets, where stochastic network calculus is used to characterize service guarantees under uncertainty and traffic bursts [15].

Caching and computation are also frequently coupled in heterogeneous IoT and DT settings. Adaptive caching schemes have been studied to jointly optimize delay and energy, indicating that storage decisions can reduce computation and transmission overheads in DT IoT services [16]. Similarly, joint computation offloading and caching has been studied in multi-layer MEC resource management, showing that coordinated handling of compute and storage can meaningfully improve efficiency when user demands and content reuse are non-uniform [17]. While caching is not the main control knob in our current model, these results reinforce a key insight: resource allocation across communication, computation, and storage should be coordinated, otherwise optimizing one dimension can shift burden and cost to another.

Compared with these studies, our formulation emphasizes a multi-edge, multi-region viewpoint where (i) traffic varies across space and time, (ii) edge resources are provisioned through slicing decisions at a longer time scale, and (iii) short-term offloading and allocation decisions operate within the sliced resource budget while optimizing profit-oriented metrics. This differs from approaches that focus mainly on one layer (e.g., radio assistance [11]) or one metric dimension (for example, AoI satisfaction [13], [14]) without explicitly modeling a two-time-scale coupling between slicing and offloading decisions.

### *B. Digital Twin–Enabled Edge Networks for Intelligent Offloading*

Digital twins provide a structured way to mirror and predict physical processes, networks, or services, enabling proactive control rather than purely reactive scheduling. In DT-enabled edge networks, the twin can capture state evolution, predict future demand, and support more robust offloading decisions under uncertainty. Several DT-focused works have advanced the design space of offloading and edge selection. DT-aided intelligent offloading with edge selection has been studied to improve decision quality through prediction and context-aware selection among candidate edges [18]. UAV-aided vehicular edge computing further highlights DT value in dynamic environments, where mobility and intermittent connectivity make static policies ineffective; DT-enabled task offloading frameworks help adapt dispatch to fast-changing conditions [19]. DT models have also been used to reduce offloading latency in 6G-oriented DT edge networks, emphasizing ultra-low-latency control loops and the need for predictive decision support [20]. These efforts collectively suggest that DT-assisted systems benefit most when prediction and control are tightly coupled, especially under mobility and non-stationary traffic.

Another important set of works considers where DT functions should reside and how they should be transferred across infrastructure. Adaptive digital twin placement and transfer in wireless computing power networks shows that a DT itself can be migrated or re-located to match computing power distribution and service demands [21]. Likewise, adaptive edge association for wireless DT networks in 6G highlights the need to couple association with DT-driven dynamics [22]. These studies motivate architectures that connect prediction, association, and provisioning as parts of a coherent loop. From a distributed systems perspective, related ideas also appear in distributed capacity optimisation and resource allocation frameworks, where decentralized decision making is used to improve scalability and capacity utilization under heterogeneous resources and dynamic demands [23]. Our work aligns with these motivations, but focuses specifically on using traffic cognition to drive a resource slicing + offloading pipeline with clear operational constraints and cost structure.

### *C. Federated Learning, Blockchain, and Incentives for DT Edge Networks*

As DT-enabled edge systems scale, centralized learning and data sharing become difficult due to privacy, bandwidth limits, and administrative boundaries. This has led to extensive work on federated learning (FL) and related coordination mechanisms in DT edge environments. Communication-efficient FL for DT edge networks in industrial IoT demonstrates how model update design can reduce overhead, which is essential when edge links are constrained [24]. FL has also been integrated with blockchain to improve trust, traceability, and integrity of distributed training. Cooperative FL and model update verification in blockchain-empowered DT edge networks shows how verification and coordination can be strengthened under adversarial risk [25]. Low-latency FL and blockchain for edge association in DT-empowered 6G networks further connects learning with control decisions, emphasizing that learning outcomes can drive association and resource control [26]. Lightweight DT and FL with distributed incentive mechanisms in air–ground 6G networks underscores that incentives matter: when devices spend energy and bandwidth, they need clear reasons to participate [27]. These works collectively show that privacy, security, overhead, and incentives must be treated as first-class design constraints.

Resource efficiency is also central. Resource-efficient adaptive FL in DT-enabled IIoT proposes adaptive strategies that allocate learning resources based on context and benefit [28]. Secrecy-driven energy minimization in FL-assisted marine DT networks shows how secrecy constraints and energy objectives reshape training and network decisions [29]. Adaptive FL for DT-driven industrial IoT has been studied to address device heterogeneity and non-stationary data [30]. Optimization of FL with deep reinforcement learning for DT-empowered industrial IoT indicates that RL can tune FL control variables for better convergence under system constraints [31]. Finally, personalization has become important in heterogeneous industrial IoT, where one global model may not fit all devices; communication-efficient personalized FL for DT in heterogeneous IIoT addresses this gap by tailoring models to device context [32]. Complementary aggregation strategies such as fine-tuning and head aggregation further improve performance under non-IID data [33].

While these studies focus on distributed learning under DT settings, our emphasis is different: the core challenge here is not primarily how to train DT models, but how to use traffic cognition and learning-based control to drive two-time-scale slicing and offloading decisions in a multi-edge system. Still, the FL literature strongly motivates privacy-aware system design: in realistic deployments, edges may not reveal sensitive task or capacity information. Our decentralized mechanism follows this same spirit by limiting information exchange while still enabling effective decisions, similar to communication-efficient learning and verification ideas [24]–[26], [28], [32], but applied to resource slicing and computation offloading rather than training alone.

### *D. Learning Foundations and Model Capacity Considerations*

Deep learning architectures underpin many predictive and decision modules used across MEC, DT, and FL systems. Foundational studies on learning multiple layers of features from data remain influential as early cornerstones that shaped representation learning and scalable training practices [32]. These foundations are relevant in modern traffic prediction and control because learning modules must generalize under non-stationary data, heterogeneous devices, and limited communication budgets—challenges that also appear in personalized and adaptive FL designs [32], [28], and in aggregation enhancements such as head aggregation [33].

### *E. Link to Earlier Adaptive Networking and Energy-Aware Design Themes*

It is also useful to relate today's traffic-cognitive edge optimization to earlier adaptive networking research. In MANET settings, efficient routing protocols have been proposed to discover optimum paths under mobility and dynamic topology changes, highlighting how quickly a network's "best" decision can shift as conditions evolve [35]. In resource-constrained sensor deployments, opportunistic routing approaches have been designed to reduce power consumption while still maintaining effective delivery, reinforcing the importance of cost-aware decision making in constrained networks [36]. Although these works focus on routing rather than MEC, the shared theme is clear: adaptive control under uncertainty and resource limits is essential. Our work translates this principle into a multi-edge context, where the "route choice" analogue is the selection of offloading targets and slice allocations under time-varying traffic and constrained edge capacity.

### *F. Summary of Gaps and Our Contribution Relative to Existing Literature*

The reviewed DT-edge and FL literature shows strong progress in prediction, learning efficiency, and trust/verification [18], [33], but gaps remain for multi-edge systems operated with slicing procurement and profit/cost structure. First, two-time-scale coupling is often implicit; many works focus on per-slot offloading, association, or DT migration without explicitly modeling long-term provisioning that constrains short-term actions [20]–[22]. Second, platform-oriented profit/cost objectives are less emphasized compared to QoS, delay, or energy goals [11], [16], [29]. Third, privacy limits and partial information require system-level decision designs, not only training-level protections [24]–[27], [32]. By addressing these gaps, our work complements existing DT-edge and FL research while aligning with broader adaptive and cost-aware networking themes seen in QoS-aware HetNet optimization [12], live streaming guarantees [15], distributed resource coordination [23], and classic adaptive routing and energy-aware forwarding [35]-[38].

## III. SYSTEM MODEL AND PROBLEM FORMULATION

Fig 1 illustrates the proposed computation offloading model towards traffic-cognitive network slicing in a multi-edge system. The system contains several edge regions, and the Base Station (BS) and edge servers in each edge region offer communication and computing resources for processing the offloaded tasks from the intelligent applications of users in the region, respectively. Users are randomly distributed within the communication coverage of each BS, and the ESP deploys services to edge servers in all regions to achieve service coverage for all users in the system. In each edge region, the ESP first sends slicing requests with demanded resources to the InP and makes payment. Next, the ESP deploys the offloading service on the edge server and manages the owned resources and system states. Finally, through paying fees, users can access the offloading service for processing their tasks.

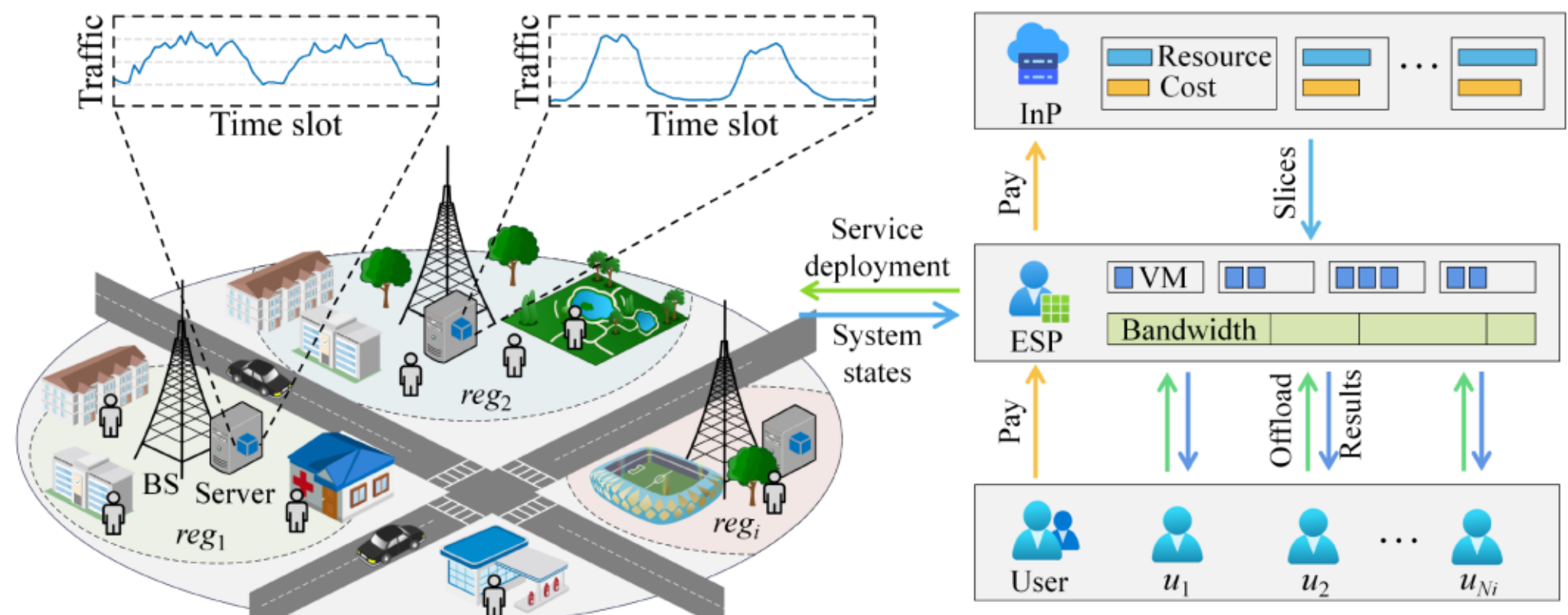

Fig 1. The proposed computation offloading model towards traffic-cognitive network slicing in a multi-edge system.

Specifically, the edge regions in the system are defined as the set $Re\,g = \{reg_1, reg_2, ..., reg_i, ..., reg|Re\,g\}$, and the users within the coverage of the region $reg_i$ are denoted as the set $U_i = \{u_i, 1, u_i, 2, ..., u_i, j, ..., u_i, , N_i\}$, where $N_i$ is the number of users in $reg_i$ he communication resources of BSs and the computing resources of edge servers are provided by bandwidths and virtual machines (VMs), respectively. Due to user mobility and request randomness, the number of users sending offloading in different regions and time-slots experiences fluctuations, causing inconsistent spatio-temporal distributions of service demands.

To improve the efficiency of resource utilization and ESP profits, the ESP analyzes recent user traffic and their resource demands in each region at regular intervals, and then makes slice resource adjustments. To avoid QoS degradation caused by frequent slice adjustments, we adopt two scales of time-slots to cope with problems with different dimensions. Specifically, a long time-slot is denoted ash, where h ∈ {1, 2, ..., H}. At the start of h, the ESP evaluates the required slice resources in each region based on the information of user traffic and offloaded tasks collected in historical time-slots and sends slicing requests to the InP. Meanwhile, the long time-slot h is divided into several short time-slots, and each is denoted as t, where t ∈ {1, 2, ..., T}. At the start of *t*, users upload their offloading requests to the ESP. The ESP assesses resource demands and user priorities and then makes proper decisions on computation offloading and resource allocation. At the end of *t*, the ESP collects the information of user traffic and task completion status in each region for subsequent slice adjustments.

### 3.1 Communication and Computation Models

The task of the user $u_{i,j}$ is clarified as a 4-tuple $< d_{i,j}, \eta_{i,j}, \rho_{i,j}, l_{i,j} >$, which are the data size, computing density, priority of $u_{i,j}$ and distance between $u_{i,j}$ and the BS, respectively. $\rho_{i,j}$ indicates the service level. More revenues can be obtained if the higher-priority tasks are completed.

When $u_{i,j}$ offloads its task to the edge server for execu-tion, the input data should first be uploaded. The bandwidth allocated to $u_{i,j}$ is denoted as $b_{i,j}^{up}$ referring to Shannon's theorem [27], the rate of task uploading is

$$r_{i,j} = b_{i,j}^{up}\, log_2(1 + \frac{pg_{i,j}}{\sigma^2})(1)$$

Where $p, \sigma^2$ and $g_{i,j} = \beta_0 l_{i,j}^{-\theta}$ re the upload power, noise power, and channel power gain between $u_{i,j}$ and the BS, respectively.

Therefore, the task uploading time is defined as

$$T_{i,j}^{up} = \frac{d_{i,j}}{r_{i,j}} (2)$$

Once a task is uploaded to the BS, the ESP will schedule the task to a specific VM for execution. A VM may execute multiple tasks simultaneously, and thus it maintains a task waiting queue. When a task is scheduled to $VM_m$ the task queuing time is defined as

$$T_{i,j}^{que} = \sum_{k=1}^{|Q_m|} \frac{d_{i,k}\eta_{i,k}}{f^{ege}}, (3)$$

Where $Q_m$ represents the task waiting queue that already exists when a task arrives at $VM_m$, and $f^{edga}$ s the com-puting frequency of $VM_m$.
Therefore, the task execution time is defined as

$$T_{i,j}^{exe} = \frac{d_{i,j}\eta_{i,j}}{f^{edga}} (4)$$

After completing tasks, the results will be returned to users. Since the output data is much smaller than the input data, the result returning time is typically ignorable [28]. Through integrating the communication and computation models, the task completion time can be calculated by

$$T_{i,j}^{total} = T_{i,j}^{up} + T_{i,j}^{que} + T_{i.j}^{exe} (5)$$

**3.2 Profit Model**

When assessing ESP profits, both the revenues and costs of processing tasks should be considered. The ESP obtains revenues from users by offering services. If user tasks are completed within the maximum tolerable delay $T^{max}$, the ESP will obtain the revenue Φ. Otherwise, no revenue will be obtained. Hence, the revenue obtained from $u_{i,j}$ within t can be described as

$$V_{i,j}^{t} = \begin{cases} \Phi, T_{i,j}^{total} \leq T^{max} \\ 0, otherwise \end{cases}$$

Further, the ESP will obtain different revenues if it completes tasks with diverse priorities. Therefore, the total revenues obtained in all regions within h is defined as

$$R^h = \sum_{t=1}^{T} \sum_{i=1}^{|Reg|} \sum_{i=1}^{N_i} V_{i,j}^{t}\rho_{i,j} (7)$$

Meanwhile, the ESP will make payment for the rented resources in each region. The available bandwidths sold by nP in $reg_i$ are defined as $B_i = \left\{b_i^1, b_i^2, \ldots, b_i^{|B_i|}\right\}$ with the costs of $\left\{\zeta_i^1, \zeta_i^2, \ldots, \zeta_i^{|B_i|}\right\}$. The available numbers of VMs sold by InP in region are defined as $V_i =$ $\left\{v_i^1, v_i^2, \ldots, v_i^{|V_i|}\right\}$ with the costs of $\left\{\zeta_i^1, \zeta_i^2, \ldots, \zeta_i^{|V_i|}\right\}$ The bandwidths and VMs rented by ESP in the $reg_i$ are presented as

$$B_{esp}^i = \sum_{k=1}^{|B_i|} \alpha_i^k b_i^k, V_{esp}^i = \sum_{k=1}^{|V_i|} \beta_i^k v_i^k, (8)$$

Where $\alpha_i^k \in \{0,1\}$ and $\beta_i^k \in \{0,1\}$ are the renting decisions of the ESP for bandwidths and VMs in region, respectively. The total bandwidths and VMs rented by ESP in all regions at time-slot h are

$$B_{esp}^h = \sum_{i=1}^{|Re\,g|} B_{esp}^i, V_{esp}^h = \sum_{i=1}^{|Re\,g|} V_{esp}^i (9)$$

Thus, the costs of renting resources within h are

$$C^h = \sum_{i=1}^{|Re\,g|} \sum_{k=1}^{|B_i|} \alpha_i^k \zeta_i^k + \sum_{i=1}^{|Re\,g|} \sum_{k=1}^{|V_i|} \beta_i^k \xi_i^k \ (10)$$

### 3.3 Problem Formulation

Our objective is to maximize long-term ESP profits, and thus the optimization problem is formulated as

$$P1: max$$

$$\alpha, \beta, b, x \sum_{h=1}^{H} (R^h - C^h)\ (11)$$

$$s.t C_1: \alpha_i^k, \beta_i^k \in \{0,1\}, \forall i, \forall k,$$

$$C_2: \sum_{k=1}^{|B_i|} \alpha_i^k = 1, \forall i,$$

$$C_3: \sum_{k=1}^{|V_i|} \beta_i^k = 1, \forall i,$$

$$C_4: \sum_{k=1}^{|N_i|} b_{i,j}^{up} \leq B_{esp}^i, \forall i,$$

$$C_5: \sum_{k=1}^{|N_i|} f^{edge} \leq V_{esp}^i, \forall i,$$

where α and β represent the bandwidth and VM renting decisions of ESP in all regions, respectively. b and x represent the bandwidth and VM allocation decisions in all regions, respectively. C1 indicates that ESP can only decide whether to fully rent a type of resource. C2 and C3 indicate that the ESP can only rent a type of bandwidth and VM. C4 and C5 indicate that the allocated bandwidths and VMs cannot exceed the resources rented by ESP.

**Theorem 1. P1 is an NP-hard problem.**

**Proof.** We aim to prove a special case of P 1 is equivalent to the Maximum Budget Coverage Problem (BMCP) that is NP-hard [29]. In BMCP, there is a set $E = \{e_1, e_1, \dots, e_n\}$, where each element owns specific value and cost. The objective of the BMCP is to find the subset $E' \subseteq E$ hat can maximize total values while meeting the cost constraint.

In the edge region $reg_i$, when $\alpha, \beta$ and b in P 1 are fixed, the offloading requests can be deemed as the elements of E, where the allocated VMs and the revenues from completing tasks are mapped to costs and values, respectively. Thus, this special case of P 1 is defined as

$$max \sum_{j=1}^{N_i} v_{i,j}^t \rho_{i,j} \quad (12)$$

$$s.t. \sum_{j=1}^{N_i} f^{edge} \leq V_{esp}^i$$

It can be found that this special case is equivalent to the NP-hard BMCP. By extending the above special case to a multi-edge system, we can derive that P 1 is NP-hard. To enhance QoS and ESP profits, it is essential to properly adjust slice resources during the offloading process. The increase or decrease of user traffic and slice resources might lead to significant changes in problem space. Moreover, the problems of network slicing and computation offloading belong to different time scales, raising the problem-solving difficulty. To relieve this issue, we decouple P 1 into two sub-problems and formulate them as follows.

P 1.1: This sub-problem is to minimize ESP resource costs in long time-slots by making slice adjustments while meeting user demands, which is defined as

$$P1.1: min \sum_{h=1}^{H} C^h \quad (13)$$

$$\alpha, \beta$$

$$s.t C1 - C3,$$

$$C6: T_{i,j}^{total} \leq T^{max}$$

## IV. THE PROPOSED SliceOff

In this section, we first present an overview of the SliceOff. Next, we describe the two core components of the SliceOff in detail and provide rigorous theoretical proofs for the effectiveness. Finally, we analyze the complexity of the SliceOff.

### 4.1. Overview of the SliceOff

The proposed SliceOff aims to improve ESP profits by jointly solving the sub-problems of network slicing (P 1.1) and computation offloading (P 1.2). For P 1.1, linear programming and random rounding are adopted to obtain the optimal slicing scheme. To enhance the cognition ability for fluctuating user traffic, a self-attention-based traffic prediction model is devised to support adaptive slice adjustments. For P 1.2, we develop an improved DRL method, where a dual-distillation mechanism is designed to explore the optimal policy from different environmental perspectives to enhance the learning efficiency and adaptability of DRL agents in huge problem spaces.

Fig. 2 illustrates the overview of the SliceOff, where the main workflow is outlined in Algorithm 1. For each long time-slot, Algorithm 2 is first called to obtain the renting decisions of bandwidth and VM of each region (i.e, $\alpha_i^k$ and $\beta_i^k$) and then perform slice adjustments and calculate resource costs based on total renting resource (i.e, $\beta_{esp}^h$ and $V_{esp}^h$) Lines 2~3). At the start of each short time-slot, offloading requests are first sent to the ESP and the task information is uploaded to the BS (Line 5). Next, Algorithm 3 is called to generate bandwidth allocation and offloading decisions (i.e, $b^t$ and $x^t$) of each region. Then, the offloaded tasks are executed on edge servers according to $b^t$ and $x^t$ and the results are returned after task completion (Lines 6~7). At the end of each short time-slot, the ESP calculates the revenues based on task completion time and user priority and collects the state information of each region for subsequent slice resource adjustments (Lines 8 ~ 9). Finally, it goes to the next long time-slot.

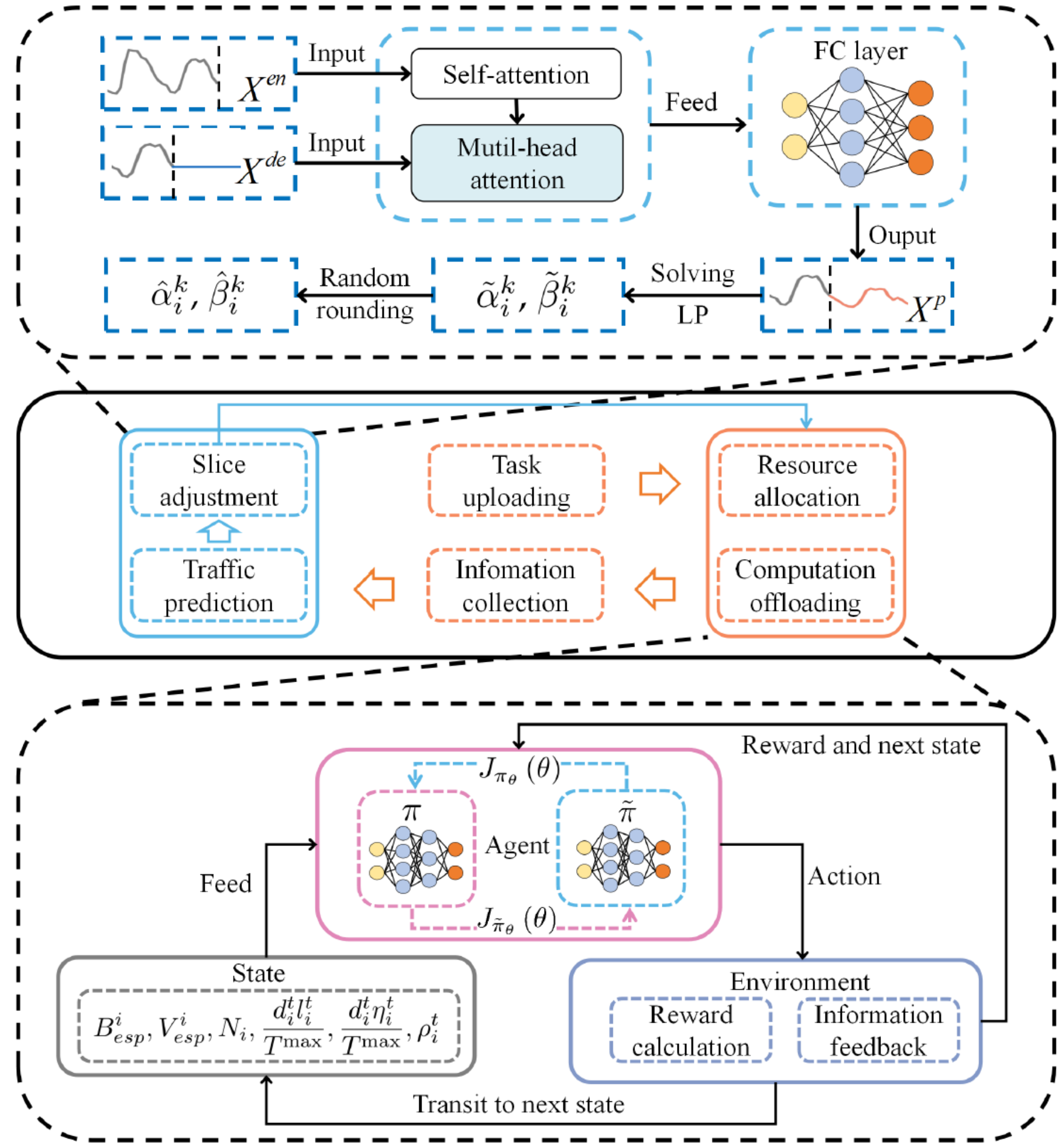

Fig. 2. Overview of the proposed SliceOff.

### *4.2 Prediction-assisted Slice Adjustment*

By evaluating user traffic and offloading demands in var-ious regions, slices can be adjusted in advance to bet-ter meet user demands and improve resource efficiency. In multi-edge environments, the traffic exhibits long-term changes (impacted by application popularity) and short-term fluctuations (impacted by user mobility), and mean-while, different edge regions may have various traffic pat-terns. These factors pose significant challenges to traffic prediction. Compared to classic Recurrent Neural Networks (RNNs) and Convolutional Neural Networks (CNNs), the emerging Transformer [30] introduces a self-attention mech-anism to capture the temporal dependencies in the se-quence, avoiding the problem of gradient explosion or vanishing. Considering its excellent ability, we design a self-attention-based method to predict user traffic in different regions.

**Algorithm 1:** The proposed SliceOff

**Input:** B, V
**Output:** Slicing and offloading policies
1 **for** h = 1, 2,..., H **do**
2 | Call **Algorithm** 2 to obtain $\alpha_i^k$ and $\beta_i^k$
3 | Perform slice adjustments and calculate resource costs based on $B_{esp}^h$ and $F_{esp}^h$
4 | **for** t = 1, 2, ..., T do
5 | Send offloading requests to the ESP;
6 | Call Algorithm 3 to generate $b^t$ and $x^t$
7 | Execute tasks and return results;
8 | Calculate ESP revenues;
9 | Collect the state information of each region;

Moreover, service demands commonly depend on the services provided by the ESP, which can be estimated by analyzing historical task attributes and system loads. Specifically, we adopt a linear programming (LP) solver to obtain optimal slicing resources based on traffic prediction and service demands analysis. Considering that P 1.1 is constrained by C1 (i.e, $\alpha_i^k, \beta_i^k \in \{0,1\}$) and the solver outputs real solutions, we utilize random rounding to round real solutions to feasible ones for making slice-adjustment decisions. The key steps of the proposed adaptive slice adjustment method are given in Algorithm 2.

**Algorithm 2:** Prediction-assisted slice adjustment

**Input:** Historical system states, B, V
**Output:** $\alpha_i^k, \beta_i^k$
1 Construct input vectors using historical □user traffic:
$X^{en} \leftarrow X^{his}, X^{de} \leftarrow Concat(X^{cur}, X^0)$;
2 Output of encoder: $H_1 = Encoder(X^{en})$;
3 Output of decoder: $H_2 = Decoder(H_1, X^{de})$;
4 Predict future traffic: $X^p = MLP(H_2)$;
5 Convert C6 of P 1.1 into Eq. (17);
6 Construct LP solver by relaxing C1 to $\alpha_i^k$ and $\beta_i^k \in [0,1]$;
7 Obtain the optimal $\tilde{\alpha}_i^k$ and $\tilde{b}_i^k$;
8 **for** i = 1, 2, ..., |Reg| **do**
9 | **for** k = 1, 2, ..., $|B_i|$ **do**
10 | Round $\tilde{\alpha}_i^k = 1$ or $\tilde{\alpha}_i^k = 0$ according to $\tilde{\alpha}_i^k$
11 | **for** k = 1, 2, ..., $|V_i|$ **do**
12 | Round $\tilde{\beta}_i^k = 1$ or $\tilde{\beta}_i^k = 0$ according to $\tilde{\beta}_i^k$

Step 1: Predict future traffic. The user traffic in historical system states is first used to construct the input vectors of the encoder and decoder (Line 1). $X^{his}$ ndicates the histor-ical traffic of all regions. $X^{cur}$ ndicates the traffic collected in the current prediction window that contains the traffic from previous long time-slots. To avoid the high complexity caused by calculating the attention weights of all historical time-slots, we adopt a probsparse self-attention [31] in the encoder and design a self-attention distillation between layers to reduce network overheads (Line 2). Specifically, the feature extraction from j-th to (j + 1)-th layers is defined as

$$X_{j+1}^{en} = MaxPool(ELU(Conv1d([X_j^{en}]_{attention})))(15)$$

$$[\cdot]_{attention} = Soft\,max(\frac{\bar{Q}K^T}{\sqrt{d}})V, (16)$$

**Algorithm 3:** Improved DRL with dual distillation for computation offloading and resource allocation

**Input:** $B_{esp}^{i}{}', V_{esp}^{i}$

**Output:** $b^t, x^t$

1 **Initialize:** current and peer DRL agents

2 **for** epoch = 1, 2, ..., E **do**

3 Initialize state: $s_0 = env.reset()$;

4 **for** t = 1, 2, ..., T **do**

5 Explore action of computation offloading and resource allocation: $a_t = \pi(s_t|\theta^{\pi}) + N_t$;

6 Feedback rt and $s_{t+1}$ after executing $a_t$: $r_t, s_{t+1} = env.step(a_t)$;

7 Store state-transition samples in replay buffer: RB.push $(s_t, a_t, r_t, s_{t+1})$;

8 Randomly select K samples: $K * (s_t, a_t, r_t, s_{t+1}) = RB.sample(K)$;

9 Obtain $\tilde{a}_{t+1}$ at $s_{t+1}$ according to Eq. (23);

10 Calculate target Q-value $y_t$ by using $r_t$ according to Eq. (24);

11 Update $Q_1$ and $Q_2$:

$$\phi^{Q_i} \leftarrow min(y_t - Q_i(s_t, a_t));$$

12 **if** t mod 2 = 0 **then**

13 Update actor's network by Eq. (25);

14 Update actor's network with $\tilde{\pi}$ by Eq. (26);

15 Update target networks via soft update;

16 Update peer agent $\tilde{\pi}$;

Where $[\cdot]_{attention}$ indicates the probsparse self-attention, d is the dimension of $X_j^{en}$, Conv1d indicates the one-dimensional convolution, ELU is the activation function, and MaxPool indicates the maximum pooling. $\bar{Q}$ is a sparse matrix that contains the Top-u queries, making the prob-sparse self-attention calculating only O(log L) dot-product for each query-key lookup, where L is the length of $X^{cur}$. Next, the output of the encoder and Xcur are fed into the decoder, which consists of a multi-head probsparse self-attention and a masked multi-head attention (Line 3). Finally, the output of the decoder is fed into the Fully Connected (FC) layer, and thus the future traffic in all regions (i.e., $X^p$) can be predicted (Line 4).

### 4.3 Improved DRL with Dual Distillation for Computa-tion Offloading and Resource Allocation

When dealing with the complex problem of computation offloading and resource allocation with variable available resources, existing DRL-based methods reveal the perfor-mance bottlenecks caused by the Q-value overestimation and low exploration efficiency. To address these issues, we propose an improved DRL method with dual distillation, whose main steps are described in Algorithm 3. Specifically, we introduce twin critics' networks with a delay mechanism to solve the Q-value overestimation and reduce variance. Meanwhile, inspired by the dual distillation [33], we adopt two DRL agents that conduct explorations from different en-vironmental perspectives and distill knowledge from each other to improve the learning efficiency in huge decision spaces. We consider the problem model of P 1.2 as the environment, and each DRL agent interacts with the envi-ronment to optimize its policy while distilling knowledge from the peer agent's policy. The state space, action space, and reward function are defined as follows.

- **State space.** It comprises the available bandwidth, VMs, user traffic, and task attributes in the current time-slot. To better capture demand features, we transfer the data size and computing density of tasks into the demands of uploading rate and computing frequency. Thus, the system state at t in $reg_i$ is defined as

$$s_t = \left\{B_{esp}^i, V_{esp}^i, N_i, \frac{d_i^t l_i^t}{T^{max}}, \frac{d_i^t \eta_i^t}{T^{max_i^t}}, p\right\} \quad (20)$$

Where $d_i^t, l_i^t, \eta_i^t$, and $\rho_i^t$ are vectors. For example, $d_i t = \left\{d_{i,1}^t, d_{i,2}^t, \ldots, d_{i,N_i}^t\right\}$.

- **Action space.** The DRL agent should simultaneously determine the bandwidths and VMs to be allocated, and thus the action at t is defined as

$$a_t = \{b_i^t, x_i^t\}, (21)$$

Where $b_i^t = \left\{b_{i,1}^t, b_{i,2}^t, \ldots, b_{i,N_i}^t\right\}$ indicates the propor-tion of bandwidths, and the bandwidth allocated to $u_{i,j}$ is $b_{i,j}^t B_{esp}^i$. $x_i^t = \left\{x_{i,1}^t, x_{i,2}^t, \ldots, x_{i,N_i}^t\right\}$ indica es the VM index, and the VM index allocated to $u_{i,j}$ is $\left|x_{i,j}^t V_{esp}^i\right|$.

- **Reward function.** The objective of solving P 1.2 is to maximize cumulative ESP revenues, and thus the reward function is defined as

$$r_t = \sum_{j=1}^{N_i} v_{i,j}^t \, \rho_{i,j}. (22)$$

In Algorithm 3, we first initialize the current and peer DRL agents (Line 1), where each agent consists of online networks (two critics' networks $Q_1$ and actor's network π) and target networks ($Q_1{}'$, $Q_2{}'$ and $\pi'$) Different from classic DRL that directly employs the maximized Q-value, the proposed method adopts two independent critics' networks to fit the Q-value function. This design alleviates Q-value overestimation and avoids getting stuck in the sub-optimum due to undeserved cumulative errors. For each training epoch, the environment is first initialized for the current DRL agent (Line 3). At each short time-slot t, the state st is fed into the actor's network π, and then the DRL agent explores the action at at the current state according to π and exploration noise (Line 5). After executing compu-tation offloading and resource allocation, the environment feedbacks the immediate reward rt and next state st+1 (Line 6). Next, the state-transition samples are stored in a replay buffer (Line 7), and then K samples are randomly selected to update network parameters (Line 8). When updating the critic's network, $\tilde{a}_{t+1}$ at $s_{t+1}$ is first obtained by the target actor's network (Line 9). This process is defined as

$$\tilde{a}_{t+1} = \pi'(s_{t+1}|\theta^{\pi'}) = \varepsilon, \varepsilon \sim N(0, \sigma), (23)$$

where the network noise ε is utilized as regularization that makes similar actions hold equivalent rewards. Next, the target Q-value is calculated by using $r_t$ and comparing two critics' networks (Line 10), which is de-scribed as

$$y_t = r(s_t, a_t) + \gamma \cdot \min_{i=1,2}(Q'_{\phi_{i,j}}(s_{t+1}, \tilde{a}_{t+1})). \quad (24)$$

Then, the two critic's networks are updated (Line 11). To decrease the updating frequency of the policies with low quality, we adopt a delay mechanism to update the actor's network and target networks. If t mod 2 = 0, the actor's network will be updated by gradient ascent (Line 13). This process is defined as

$$\nabla_\theta J(\theta) = \frac{1}{K}\sum_{j=1}^{K} \nabla_\alpha Q_1(s, \alpha)|\alpha = \pi(s)\nabla_{\theta\pi\theta}(s). (25)$$

Next, the policy of the current agent is updated with the peer agent's policy $\tilde{\pi}$. However, it is hard to determine which of the two agents performs better at different states. To address this issue, we soften the loss function and introduce a weighted loss function to update the policy (Line 14). Specifically, the distillation objective function is defined as

$$J_{\tilde{\pi}}(\theta) = \mathrm{E}_{s \sim \tilde{\pi}}\left[\| \pi(s) - \tilde{\pi}(s) \|_2^2 \exp(\alpha \xi^{\tilde{\pi}}(s))\right], \quad (26)$$

where exp ($\alpha_{\tilde{\pi}}(s)$) is the confidence score, and α controls the confidence level that depends on the accuracy of value-function estimation. $\xi^{\tilde{\pi}}(s)$ indicates the advantage value of $\tilde{\pi}$ compared to π at s, which is defined as

$$\xi^{\tilde{\pi}}(s) = V^{\tilde{\pi}}(s) - V^{\pi}(s), (27)$$

where the larger $\xi^{\tilde{\pi}}(s)$ indicates the loss function is more conducive for the current policy to learn from the peer one. Otherwise, it is more inclined to remain unchanged.

Next, the target networks are updated via soft update (Line 15), and thus the critic's network is updated more fre-quently than the actor's and target networks. Unlike directly updating all networks, this manner reduces cumulative errors and improves training stability. Finally, the algorithm is continuously running to update the peer agent's policy (Line 16). During this process, each agent is updated ac-cording to its reward function and distillation loss function, achieving its own update while learning useful knowledge from the peer agent to enhance itself.

*Theorem 3*. Dual distillation between the current and peer agents supports achieving a more optimized hybrid policy.

**Proof**. A hybrid policy consists of the current and peer agents' policies that are chosen according to the relative advantage value, and it is defined as

$$\pi^*(s) = \begin{cases} \pi(s), \xi^{\tilde{\pi}}(s) > 0 \\ \tilde{\pi}(s), otherwise \end{cases}. \quad (28)$$

Therefore, it can be guaranteed that $\pi^*$ is a more opti-mized policy than π and $\tilde{\pi}$ (i.e, $\nabla s, V^{\pi^*}(s) \geq V^{\pi}(s)$ and $V^{\pi^*}(s)$). V π (s) ≥ V π˜ (s)). Next, we consider a simple form of the objective function, which is defined as

$$J'_{\tilde{\pi}}(\theta) = \mathrm{E}_{s \sim \tilde{\pi}}\left[\| \pi(s) - \tilde{\pi}(s) \|^2 1(\xi^{\tilde{\pi}}(s) > 0)\right], \quad (29)$$

Where 1 is the indicator function. If $\xi^{\pi'}(s) > 0$, the value of 1 (·)is 1. Otherwise , this value is 0. Since $\rho_{\pi^*}$ and $\rho_{\tilde{\pi}}$ is similar, the difference between $\rho_{\tilde{\pi}}$ and $\rho_{\pi^*}$ is negligible. Thus, the proposed dual-distillation process can be described as

$$\begin{aligned}
&E_{s\sim\tilde{\pi}}\left[\| \pi(s) - \tilde{\pi}(s) \|^2 \; 1(\xi^{\tilde{\pi}}(s) > 0)\right] \\
&= \sum_{s\sim\rho\tilde{\pi};\xi^{\tilde{\pi}}(s)>0} \| D(\pi(s) - \tilde{\pi}(s)) \|^2 + \sum_{s\sim\rho\tilde{\pi};\xi^{\tilde{\pi}}(s)\leq 0} \| D(\pi(s) - \pi(s)) \|^2 \\
&= \sum_{s\sim\rho\pi^*;\xi^{\tilde{\pi}}(s)>0} \| D(\pi(s) - \tilde{\pi}(s)) \|^2 + \sum_{s\sim\rho\pi^*;\xi^{\tilde{\pi}}(s)\leq 0} \| D(\pi(s) - \pi(s)) \|^2 \quad (30) \\
&= \sum_{s\sim\rho_{\pi^*}} \| \pi(s) = \pi^*(s) \|^2 \\
&= E_{s\sim\pi^*}[\| \pi(s) - \pi^*(s) \|^2].
\end{aligned}$$

Based on the above proof, we can derive that the knowl-edge of π˜ will be transferred to π if the advantage value of $\tilde{\pi}$ is positive. Otherwise, the update of the current policy will ignore the dual-distillation process. Therefore, the dual-distillation enables a more optimized hybrid policy. Further, to reduce the errors of value-function estimation caused by neural networks, we replace 1 ($\xi^{\tilde{\pi}}(s)$) with exp ($\xi^{\tilde{\pi}}(s)$) and use α to control the confidence level for obtaining the objective function.

## V. Performance of Centralized Mechanism

We begin by evaluating the performance of our centralized resource allocation and task offloading mechanism. Figure 3 presents the platform and edge servers (ESs) revenue, the number of offloaded tasks, and the resource utilization efficiency of ESs under different algorithms. The comparison includes our proposed mechanism, an auction-based algorithm, a max-transaction algorithm, and a greedy approach. This analysis considers a scenario with 3 ESs and 10 applications. From the results in Figure 3, our mechanism demonstrates significant advantages, achieving the maximum revenue for the platform. Specifically, it outperforms the auction-based algorithm, max transaction, and greedy algorithm by 33%, 81%, and 131%, respectively. Additionally, our mechanism enables a higher number of tasks to be offloaded to ESs while maintaining superior resource utilization efficiency.

Interestingly, although the auction-based algorithm offloads a greater number of tasks and the max-transaction algorithm achieves higher resource utilization efficiency, both yield lower revenue for the platform compared to our mechanism. This finding highlights an essential insight: maximizing the number of offloaded tasks or overall social welfare does not necessarily result in optimal revenue for the platform. Furthermore, Figure 3(c) reveals that the revenue generated for ESs under our mechanism is the lowest among the algorithms analyzed. A closer examination of our model clarifies this outcome.

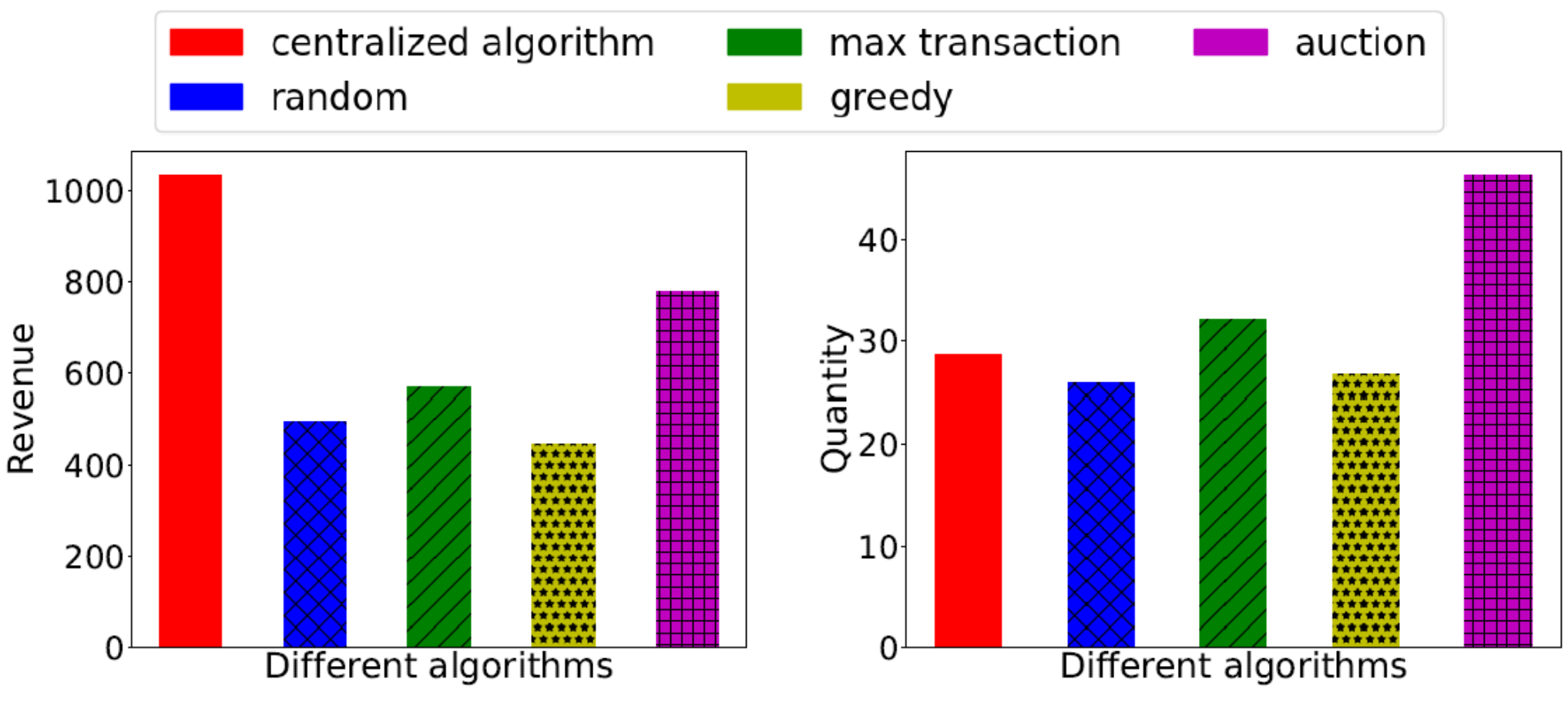

(a) Revenue of the platform (b) Offloaded Tasks

Fig. (3.a), (3,b): Performance of different mechanisms

Since the primary goal of our mechanism is to maximize the platform's revenue, it achieves this by retaining a significant portion of the payment for the platform while providing minimal rewards to ESs, as long as they remain willing to offer their services. In essence, our approach prioritizes maximizing platform profits, even at the expense of the backend ESs' revenue.

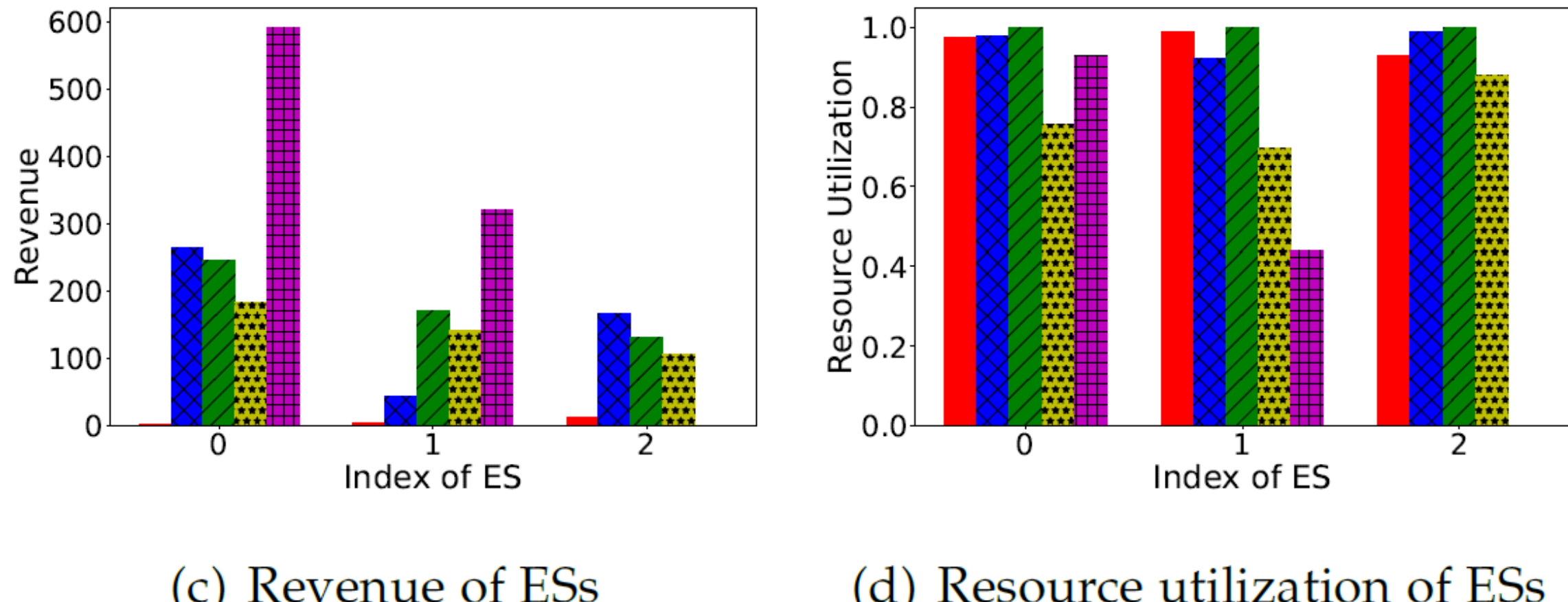

(c) Revenue of ESs (d) Resource utilization of ESs

Fig. (3c), (3d): Performance of different mechanisms

This observation underscores a fundamental trade-off in the design of incentive mechanisms for crowdsourcing-like MEC systems. While our mechanism achieves superior platform revenue, it does so by strategically allocating resources and setting rewards in a way that discourages excessive payouts to ESs. This aligns with the Stackelberg game structure, where the platform as the leader optimizes its strategy based on the rational responses of the ESs. The relatively low revenue for ESs can be attributed to the system's ability to leverage the competitive environment, ensuring that even minimal incentives are sufficient to secure participation. Moreover, this outcome reflects the inherent flexibility of our model, which allows task allocation to be adjusted dynamically in response to changing workloads and resource constraints, resulting in consistently high utilization of ES resources. Additionally, the design ensures compliance with QoS requirements by prioritizing tasks that contribute most significantly to platform revenue. By carefully balancing reward allocation, resource efficiency, and task prioritization, our mechanism not only outperforms traditional approaches but also provides a robust framework for adapting to diverse operational conditions in edge computing environments. These insights reveal the potential for tailoring similar frameworks to optimize revenue in other decentralized, multi-agent systems.

## VI. Conclusion

This study addressed the critical challenge of developing efficient strategies for resource allocation and incentive design in crowdsourcing-inspired edge computing systems. By focusing on a scenario where application providers make prepayments for offloading tasks, we developed a framework that balances the diverse interests of the platform and edge servers (ESs). To achieve this, we modeled the interactions between the stakeholders using a Stackelberg game and introduced innovative solutions tailored to different system constraints. The centralized approach leverages Bayesian optimization to streamline resource distribution and task offloading while prioritizing platform revenue. Meanwhile, the decentralized mechanism provides a robust alternative that preserves participant privacy by relying on aggregated data and local decision-making processes. This ensures system functionality even when sensitive information cannot be shared. Our numerical experiments demonstrated the effectiveness of these solutions, revealing that the centralized mechanism maximizes platform profitability, while the decentralized approach enhances ESs' individual benefits under privacy constraints. These results highlight the potential for integrating economic models with advanced computational techniques to create scalable, secure, and adaptable edge computing systems for the future.